\documentclass[showpacs,showkeys,aps,twocolumn]{revtex4}
\usepackage{bm}
\usepackage{amsmath}
\usepackage{graphicx}
\usepackage{subfigure}
\usepackage[usenames,dvipsnames]{color}
\definecolor{darkblue}{RGB}{0,0,196}
\usepackage[colorlinks=true,linkcolor=darkblue,citecolor=darkblue,urlcolor=darkblue]{hyperref}
\usepackage{setspace}
\usepackage{hyperref}
\usepackage{xcolor}
\hypersetup{
  colorlinks   = true, 
  urlcolor     = red, 
  linkcolor    = blue, 
  citecolor   = blue 
}
\usepackage{footmisc}
\usepackage[makeroom]{cancel}
\usepackage{comment}
\usepackage{lineno}
\def\be{\begin{equation}}
\def\ee{\end{equation}}
\def\ba{\begin{eqnarray}}
\def\ea{\end{eqnarray}}
\usepackage{graphicx}
\usepackage{amsmath,bbm}
\usepackage{amssymb,bm}
\usepackage{yfonts}
\usepackage{multirow}

\begin{document}
\title{Possible early universe signals in proton collisions at the Large Hadron Collider}
\author{Raghunath Sahoo$^{1,2,}$\footnote{Corresponding author: Raghunath.Sahoo@cern.ch }}
\author{Tapan Kumar Nayak$^{2,3}$}
\affiliation{$^{1}$Department of Physics, Indian Institute of Technology Indore, Simrol, Indore 453552, India}
\affiliation{$^{2}$CERN, CH 1211, Geneva 23, Switzerland}
\affiliation{$^{3}$School of Physical Sciences, National Institute of Science Education and Research, HBNI,
Jatni-752050, India}

\begin{abstract}
\noindent
Our universe was born about 13.8 billion years ago from an extremely hot and dense singular point, in a process known as the Big Bang. The hot and dense matter which dominated the system within a few microseconds of its birth was in the form of a soup of elementary quarks and gluons, known as the quark-gluon plasma (QGP). Signatures compatible with the formation of the QGP matter have experimentally been observed in heavy-ion (such as Au or Pb) collisions at ultra-relativistic energies. Recently, experimental data of proton-proton (pp) collisions at the CERN Large Hadron Collider (LHC) have also shown signals resembling those of the QGP formation, which made these studies quite stimulating as to how the collision of small systems features in producing the early universe signals. In this article, we report on some of the compelling experimental results and give an account of the present understanding. We review the pp physics program at the LHC and discuss future prospects in the context of exploring the nature of the primordial matter in the early universe.\pacs{}
\end{abstract}

\date{\today}
\maketitle

\section{Introduction}
\label{intro}
The discovery of electron as the first elementary particle by J.J. Thomson in 1897 was a major milestone in our quest to explore and understand the subatomic universe. Further down the line in the year 1911, the nucleus as a centrally placed heavy object inside an atom, with protons and neutrons (collectively known as nucleons) as its constituents was probed through the famous Rutherford alpha particle scattering experiment. The structure of nucleons was further probed by the famous deep inelastic scattering of electrons on protons which led to the discovery of the substructure of protons in the Stanford Linear Accelerator Centre (SLAC), USA in the year 1968. Later experiments used muons and neutrinos to understand the detailed structure of the hadrons. It has now been understood that protons and neutrons are composed of quarks and are bound together by gluons. The gluon was discovered at the electron-position collider (PETRA) of DESY, Germany in 1979. The fact that independent existence of these quarks and gluons is not yet directly observed in experiments, is supported by their underlying dynamics, which is known through the theory of strong interaction -- Quantum Chromodynamics (QCD). The strong interaction, one of the four fundamental interactions of nature predicts the confinement of partons (quarks and gluons) inside the cage of hadrons (the bound state of partons, e.g. proton, neutron).

However, the human curiosity of creating a matter with quarkonic degrees of freedom in the laboratory has led to a completely new field of research – search and study of the quark-gluon plasma (QGP). Incidentally, the QGP matter dominated our universe till about a few microseconds after the Big Bang. Our universe began with a Big Bang about 13.8 billion years \cite{Ref0} ago from a “point” called the singularity, which exploded and started to expand very rapidly. In the first few instants of time, between $10^{-35}$ seconds and $10^{-32}$ seconds, it underwent a period of exponential “inflation”. Until a few microseconds from the beginning (time t = 0), this hot and dense matter was in the form of QGP, consisting of deconfined (free) quarks and gluons \cite{Ref1,Ref2}. 

\begin{figure}[ht]
\includegraphics[scale=1.2]{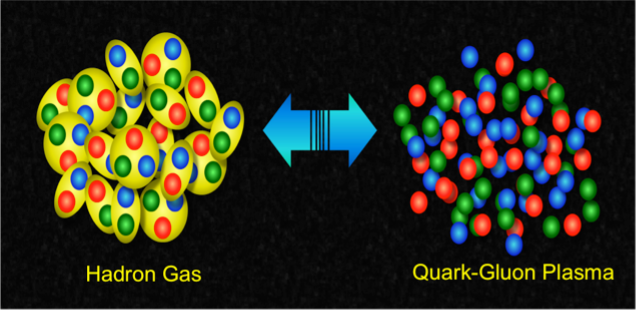}
\caption[]{(Color Online) Sketch of a system of hadron gas consisting of protons, neutrons, etc., and a system of free quarks and gluons, called the quark-gluon plasma.}
\label{fig1}
\end{figure}

Understanding the evolution of our universe during its infancy involves creating and studying the formation of QGP in the laboratory. To achieve this, first of all, we need to probe the sub-nucleonic scale of matter. For this, we take the help of the general principle of optics, where the wavelength, $\lambda$ of light (wave associated with the probe) should be less than or equal to the dimension of the object. To have a grasp on the associated energy scale, the famous de Broglie equation, $\lambda  = h/p $ ($h$ is Planck’s constant and $p$ is momentum) helps us in making an estimate that to probe the proton structure one needs energy of the order of giga-electron Volt (GeV) (note that the charge radius of a proton is around 0.877 fermi). The next step involves accelerating heavy nuclei (such as gold or lead) to high energies (GeV and tera-electron Volt (TeV)) and making head-on collisions. The collision process produces an extremely hot and dense system so as to melt the protons and neutrons to their fundamental constituents, hence a relatively large number of deconfined quarks and gluons coexist for a brief time. This is demonstrated in Fig. \ref{fig1}, where the sketch on the left shows a system consisting of protons, neutrons, etc. (called hadrons) in which quarks are bound within their boundaries and a system of free quarks and gluons on the right.

\begin{figure}[ht]
\includegraphics[scale=1.2]{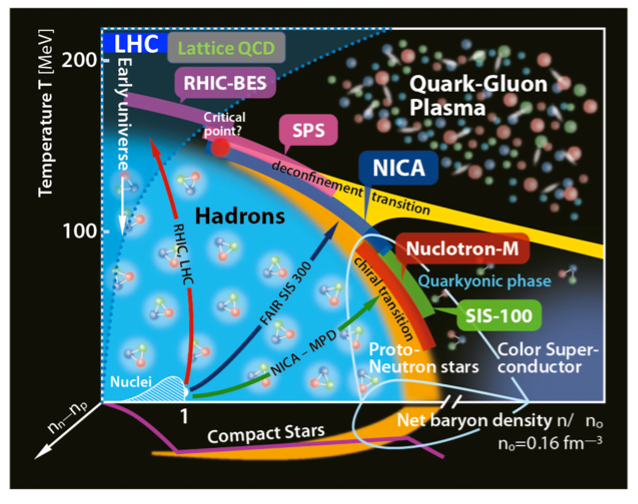}
\caption[]{(Color Online) A schematic of QCD phase diagram.}
\label{fig2}
\end{figure}

\section{Phase Diagram}
\label{section2}
The QGP state exhibits fundamentally different properties from the normal hadronic state of matter. This can be understood from the phase diagram as shown in Fig. \ref{fig2} in terms of temperature and net baryon density. At lower temperatures and finite baryon density, we have a system of hadron gas, whereas, at higher temperatures in the phase diagram, we get the QGP state. 

An early universe that started with extremely high temperature and energy density has gone through many complex dynamical processes during its spatio-temporal evolution. A high temperature and low net baryon density in the diagram represent an early universe scenario. On the other hand, a lower temperature and high net baryon density refers to astrophysical objects like neutron stars. In the domain of lower temperatures and high-baryon densities, the deconfined phase of quarks and gluons is separated by a first-order phase transition line from the hadronic matter. This line ends with a possible critical end-point (CP), after which a cross-over transition exists while moving towards higher temperatures \cite{Ref3}.

\section{QGP Signatures}
\label{section3}
Indications of the QGP formation have been observed twenty years ago in fixed-target experiments at the CERN Super Proton Synchrotron (SPS) in Geneva, Switzerland, with compelling evidence for the formation of a new state of matter at energy densities about twenty times larger than that of normal nuclear matter and temperatures 100000 times higher than in the core of the Sun. Experiments at the Relativistic Heavy-Ion Collider (RHIC) at Brookhaven National Laboratory, USA not only confirmed the observations made at SPS, but also complemented it with new insights on new observables, such as anisotropic flow and jet quenching. For the last ten years, experimental data at the LHC have opened up completely new doors to our understanding of the QGP matter.

The signatures of QGP are indeed indirect, as the lifetime of QGP is of the order of $10^{-23}$ second ($\sim$ 3 fermi/c). Most of the signals probe the nuclear medium by considering proton-proton collisions as a baseline. Some of the signatures of QGP are the observation of high energy density, high-temperature matter, enhancement of strangeness, azimuthal anisotropy, elliptic flow, collective radial expansion, suppression of quarkonia,  etc. \cite{Ref4,Ref5}. 

The main point of our discussion here is the striking observation of QGP-like signatures in proton-proton collisions at the LHC accompanied by a large number of produced particles. This warrants a relook at the use of proton-proton collisions as a baseline measurement and also to understand the formation of QGP-droplets in these collisions. The main thrust of this article is to review the QGP-like signatures in LHC proton-proton collisions.

Following the lattice QCD estimations, the critical energy density and temperature for a deconfinement transition are predicted to be, $\epsilon_c$ = 1 GeV/$\rm fm^3$ and $T_c \approx$  150-170 MeV, respectively \cite{Ref5}. To have a better realization of these numbers, the energy density of a normal nuclear matter is, $\rho_N$ = 0.17 GeV/$\rm fm^3$ and the energy density of a nucleon is, $\rho_p$ = 0.5 GeV/$\rm fm^3$ (taking proton charge radius ~ 0.877 fm). In comparison to a hadron gas, the degrees of freedom for a QGP is an order of magnitude higher, which signals the deconfinement transition \cite{Ref4}. Some of the important signatures of QGP are briefly discussed here and for completeness, one can have a look at any standard books on QGP or high energy heavy-ion collisions \cite{Ref6,Ref7}. It should also be noted here that, as we explore new energy frontiers, the particle production dynamics and the system properties evolve with energy. This makes energy-dependent modifications in some of the signatures of quark-gluon plasma.

 \section{Collisions at the LHC}
\label{section4}
\begin{figure}[ht]
\includegraphics[scale=1.2]{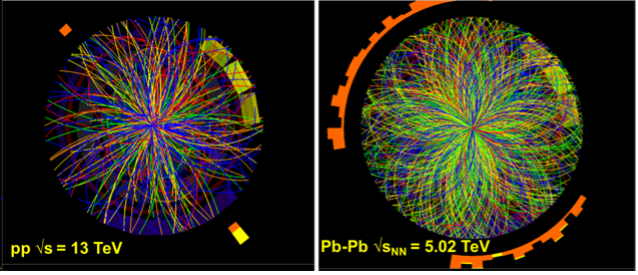}
\caption[]{(Color Online) Single event display of pp collisions at $\sqrt{s}$ = 13 TeV (left) and Pb-Pb collisions at 
$\sqrt{s_{\rm NN}}$ = 5.02 TeV (right) showing the number of charged particle tracks measured by the ALICE detector.}
\label{fig3}
\end{figure}

During the last century, science and technology have made substantial advancements in various domains. Using the cryogenic and vacuum technologies with RF accelerators, it has been possible to achieve tera-electron volt (TeV) energies in the laboratory. The Large Hadron Collider (LHC) at CERN is an example of technological excellence. After more than ten years of construction, LHC started beam operations by colliding proton on proton (pp) and lead on lead (Pb-Pb) in the year 2010. So far, the highest achieved centre-of-mass energies are 13 TeV for pp and 5.02 TeV per nucleon for Pb-Pb collisions. These high-energy collisions produce a large number of particles. Probing the hot and dense matter produced in high-energy collisions is one of the major tasks of the experiments at the LHC. ALICE experiment is devoted to the search and study of QGP in the laboratory. The other three large experiments, ATLAS, CMS, and LHCb, have excellent programs for QGP research.

Fig. \ref{fig3} shows charged particle tracks measured by the ALICE experiment for proton-proton (left) and Pb-Pb (right) collisions. When the charged particles traverse the large gas volume of the time projection chamber (TPC) of the experiment in the presence of the magnetic field, they form interesting patterns as seen in the figure. We observed that the number of produced particles in proton-proton collisions (about one hundred in a single collision event) is considerably less than Pb-Pb collisions (order of several thousand because of more number of participant nucleons and their binary collisions) at the same centre-of-mass energies. A large number of particle production is one of the characteristics of the formation of highly dense matter. A class of pp collisions where large number of particles are produced, reveals signatures similar to those of the heavier Pb-Pb collisions.

\section{QGP-like Signatures in Proton Collisions}
\label{section5}
On average, a pp collision at the LHC energies produce about 5-10 particles in the central rapidity region. However, there are certain classes of events where this number becomes much more ($\sim$ 10 times larger). What could be the reason for these high-multiplicity pp events at the LHC energies? As we know with higher energies, one probes low-Bjorken-x \footnote{Bjorken-x = $\frac{p_T}{\sqrt s}e^{-y}$, $p_T$ is the transverse momentum, and $y$ is the rapidity. This scaling variable gives the fraction of the proton momentum carried by the parton. At high energies, one probes low Bjorken-x and thus the number of quark-antiquark pairs increase.} . By this, we start seeing more sea quarks and gluons inside a nucleon (recall that there are three valence quarks inside a nucleon). With higher centre-of-mass energies available, energy per participant is possibly sufficient enough to go through inelastic interactions producing particles at the partonic level. Although the thermalization of the partonic constituents is a matter of deeper investigation, the final state high-multiplicity is an outcome of multiple partonic interactions. This aspect is taken into account in theoretical models to explain various features in experimental data. The energy density of the system produced in high energy collisions is one of the important signatures of the formation of QGP. Several phenomenological works are ongoing in that direction \cite{Ref8,Ref9}. Let us now review some of the QGP-like observations in high-multiplicity pp events for setting a stage for the search for QGP-droplets in pp collisions.
 
\subsection{Large radial flow velocity}
\begin{figure}[ht]
\includegraphics[scale=1.2]{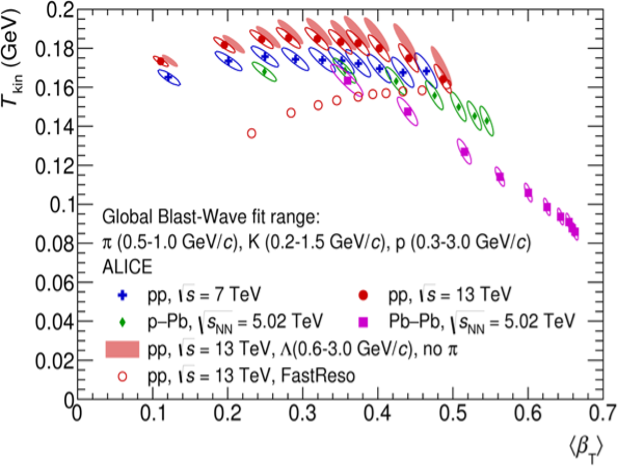}
\caption[]{(Color Online) Kinetic freeze-out temperature and radial flow velocity for pp, p-Pb, and Pb-Pb collisions at CERN LHC energies \cite{Ref17}.}
\label{fig4}
\end{figure}

One of the first and most important measurements in high-energy collisions is the transverse momentum ($p_T$) spectra of charged particles, which reveal the dynamics of the produced system, including the temperature and degree of collective expansion of the system. The $p_T$-spectra of particles produced in a multi-particle final state is expected to follow a Boltzmann-type exponential distribution with inverse-slope giving information about the temperature of the system. With higher collision energies, perturbative QCD (pQCD) processes start to make visible contributions and consequently, deviation from the usual Boltzmann type distributions have been observed. The $p_T$-spectra follow a Tsallis-Levy distribution \cite{Ref10,Ref11,Ref12,Ref13,Ref14}. Unlike the Boltzmann distribution, which has a statistical origin, the latter one is more empirical – with a combination of exponential and power-law. It should be noted here that a similar feature of the $p_T$-spectra was first conjectured by Rolf Hagedorn \cite{Ref15}. The tail of the $p_T$-spectra gets its contributions from pQCD processes, dominated by the production of jets (narrow cone of particles produced by the hadronization of quark or gluons). 

Fits of the low part of $p_T$-spectra (up to 2-3 GeV/c) of identified charged particles using the Blast-wave model yield the temperature at the kinetic freeze-out ($T_{\rm kin}$) and the average collective radial flow velocity ($<\beta_T>$) of the system. Figure \ref{fig4} shows the results of Blast-wave fits in terms of $T_{\rm kin}$ and $<\beta_T>$ for proton-proton, proton-Pb, and Pb-Pb collisions \cite{Ref16,Ref17}. Going from left to right along the x-axis, one moves from collisions yielding low to high particle multiplicity. For pp collisions at 7 TeV collision energy, this analysis yields, $T_{\rm kin} = 163\pm10$ MeV and $<\beta_T> = 0.49 \pm 0.02$ \cite{Ref17}. The high-multiplicity pp events show a high degree of collectivity, similar to those for heavy ions. This paves a way forward to look into additional signatures of QGP in high-multiplicity pp events at the LHC energies.

\subsection{Strangeness Enhancement}
\begin{figure}[ht]
\includegraphics[scale=1.2]{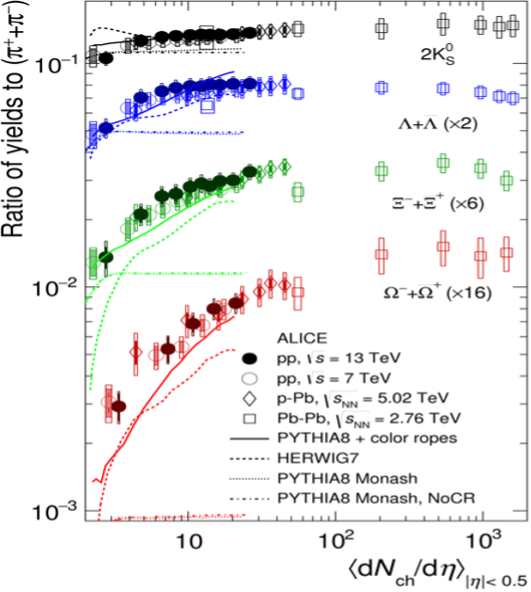}
\caption[]{(Color Online) Ratios of strange and multi-strange hadrons to pions as a function of mid-rapidity charged particle density for pp, p-Pb, and Pb-Pb collisions at CERN LHC energies. The results show heavy-ion-like strangeness enhancement in high-multiplicity pp collisions \cite{Ref16,Ref17}.}
\label{fig5}
\end{figure}

In pp or heavy-ion collisions, the colliding objects are nucleons having up ($u$) and down ($d$) quarks as constituents. However, in the final state one observes hadrons consisting of $u, d$ as well as other heavier quark flavors. Strange ($s$) quarks are the lightest among the rest of the quarks. These quarks (other than $u$ and $d$) must have been formed as a part of a possible partonic medium. In an equilibrated QGP as the temperature of the system is higher than the mass of a strange quark ($m_s \gg m_{u,d}$), strange quarks and anti-quarks can be abundantly produced through several processes leading to strangeness enhancement. These processes are flavor creation ($qq \rightarrow s\bar{s}$,  $gg \rightarrow s\bar{s}$ ), gluon splitting ($g \rightarrow s\bar{s}$ ), and flavor excitation ($gs \rightarrow gs$, $qs \rightarrow qs$). Observation of enhanced multi-strange particles in comparison to the pion yields in the final state is called strangeness enhancement and is considered as a signature of the formation of QGP. Strangeness enhancement has been experimentally observed in experiments at the CERN SPS \cite{Ref18}, RHIC \cite{Ref19}, and LHC \cite{Ref20} energies in heavy-ion collisions.

To probe into possible QGP formation, yield ratios of particles with strange and multi-strange quarks to those of pions (without strangeness content) are plotted as shown in Fig. \ref{fig5}. The results, reported by the ALICE experiment at the LHC, show that  the high-multiplicity pp collisions show similar degrees of enhancement like Pb-Pb collisions. This is an intriguing observation at the LHC energies.

\subsection{Multiparticle Ridge-like Correlations}
\begin{figure}[ht]
\includegraphics[scale=1.2]{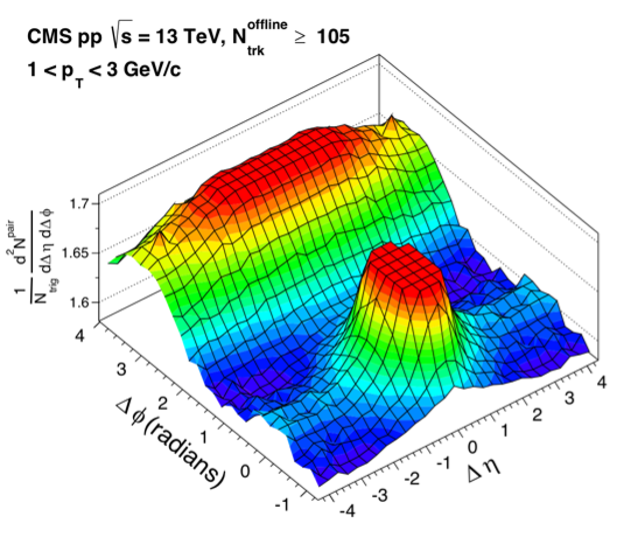}
\caption[]{(Color Online) Two-particle correlation function in high multiplicity pp collisions at $\sqrt{s}$ = 13 TeV for pairs of charged particles showing the ridge structure, with each particle within $ 1 < p_T < 3 $ GeV/c \cite{Ref24}.}
\label{fig6}
\end{figure}

The collective flow of strongly interacting matter gives rise to azimuthally collimated  long-range (large $\Delta \eta$), near-side (small $\Delta \phi$) ridge-like structure in two particle azimuthal correlations. It was first observed at RHIC in Cu-Cu \cite{Ref21} and Au-Au \cite{Ref22} collisions and later at the LHC, in Pb-Pb collisions \cite{Ref23}. The reason for the ridge formation in heavy-ion collisions is understood to be due to hydrodynamic collective expansion of strongly interacting matter, which  develops long-range (in rapidity, $|\Delta \eta| \approx 4$) correlations. Although it was not expected to be formed in proton-proton collisions, the CMS experiment at the LHC, has reported a same-side ($\Delta \phi \sim 0$) ridge in the two-particle correlations produced in high-multiplicity pp collisions as shown in Fig. \ref{fig6} \cite{Ref24}. This has opened up a new domain of theoretical investigations, which is not explained by pQCD based models like PYTHIA \cite{Ref25}. In contrast to the collisions of structured objects (such as pp, p-Pb, and Pb-Pb), the collision of point-particles (like in $e^+e^-$) show no such ridge structure as observed by the Belle experiment at KEK, Japan for $\sqrt{s}$ = 10.52 GeV \cite{Ref26}. However, it would be more interesting to have $e^+e^-$ collisions at TeV energies for a better comparison. This makes the observation of ridge in high-multiplicity pp collisions more interesting in terms of the formation of QGP droplets.

\subsection{Small system Collectivity}
\begin{figure}[ht]
\includegraphics[scale=1.1]{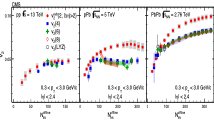}
\caption[]{(Color Online) The elliptic-flow harmonic $v_2$ as a function of particle multiplicity in pp, pPb, and PbPb collisions \cite{Ref25}. High-multiplicity pp collisions show comparable collectivity as compared to p-Pb, and Pb-Pb collisions.}
\label{fig7}
\end{figure}

The observation of an unexpected ridge-like phenomenon in high-multiplicity pp collisions gave a boost to experimental and theoretical explorations to look for more hydrodynamic behaviour in small systems. This slowly became a centre-stage research at the LHC energies to understand the origin of collectivity in small systems. The primary measurement in this direction is azimuthal asymmetry or elliptic flow harmonics of produced particles, measured from two- and multi-particle correlations. This has been measured by the CMS collaboration \cite{Ref27} 
as a function of final-state multiplicity and compared with the corresponding measurements from p–Pb and Pb–Pb collisions (Figure \ref{fig7}). The observed signals show nearly no dependence on the number of correlated particles in pp collisions like those of p-Pb and Pb-Pb collisions, providing a consistent picture for the collective nature of the medium formed in all three collision systems. It is indeed surprising to see that the observed elliptic flow, which in heavy-ion collisions proved that the matter created as {\it `perfect fluid’} showing the shear viscosity-to-entropy density ratio below the AdS/CFT conjectured minimum value of $1/4\pi$, is independent of the number of correlated particles \cite{Ref28}. This gives direct evidence of the collective origin of the observed long-range ridge. This observation got further strengthened when the famous “mass ordering” of the elliptic flow, as a perfect hydrodynamic behaviour was observed for identified particles in a range of $p_T \leq 2.5$ GeV/c in pp collisions \cite{Ref28}.  This has opened up a new direction of research in an unexpected territory to look for QGP matter formation in pp collisions.

\section{Summary and future directions}
The study of early universe signatures, in the form of deconfined quarks and gluons, has been the domain of heavy-ion collisions at relativistic energies. This stems from the fact that one can form an extended system of  QGP in heavy-ion collisions. Experiments at RHIC and LHC have confirmed the formation of QGP in these collisions after analysing several signatures. Within the last few years, several signatures of QGP, like those found in  heavy-ion collisions, have been observed in proton-proton collisions at the LHC, where the average particle multiplicities are large. These high-multiplicity events originate because of possible multiple-partonic interactions. Strangeness enhancement, formation of ridge structure in multiparticle correlations, collective hydrodynamic behaviour, higher radial flow velocity, and collison geometry driven azimuthal anisotropy are some of the observations confirming  the formation of QGP-like systems in proton-proton collisions. Although it is tempting to make a conclusion of the formation of QGP droplets in LHC pp collisions, several other observations seen in heavy-ion collisions need to be explored in small systems. Some of these include jet quenching and medium energy loss, suppression of high-$p_T$ hadrons, suppression of charmonia and bottomonia, etc. Information about the size and the nature of QGP-droplets would be of greater interest to the scientific community. 

In this direction, experiments at the LHC have entered into a new domain with detector upgrades to acquire data at a high rate with high luminosity proton-proton collisions. This will open up new avenues for accessing high multiplicity events as well as for short-lived particles, exotica, and heavy-flavors. The next LHC data-taking periods (extending to the year 2036) will be interesting for studying the details of the QGP droplets \cite{Ref29}.

The planned Future Circular Collider (FCC) \cite{Ref30} at CERN is expected to collide proton on proton at a centre-of-mass energy of 100 TeV and heavy-ion collisions at 39 TeV per nucleon. This future accelerator is planned to be of 100 Km circumference, as compared to 27 Km of the present LHC. This new energy frontier would facilitate the study of the early universe signals with a variety of new observations.

\section*{Acknowledgements}
The authors would like to thank Prof. Y.P. Viyogi, INSA Senior Scientist for his insightful comments and suggestions.


\end{document}